\documentclass[11pt,twoside]{article}

\usepackage{asp2006}
\usepackage{epsf}
\usepackage{psfig}
\usepackage{lscape}
\usepackage{epsfig,subfigure,amsmath,amssymb,paralist}

\begin{document}
\title{A 50 MHz System for GMRT}
\author{N. Udaya Shankar, K.S. Dwarakanath, Shahram Amiri, R. Somashekar,\\
B.S. Girish, Wences Laus and Arvind Nayak}
\affil{Raman Research Institute, Bangalore 560 080, India\\
{\tt\small Email: uday@rri.res.in}}

\begin{abstract} 
This paper describes a 50~MHz system being developed for GMRT to provide
imaging capability in the frequency range 30-90~MHz. Due to its larger
collecting area and higher antenna efficiency, the low frequency GMRT
system will be several times more sensitive than the present 74~MHz VLA 
system and is likely to remain a competitive instrument in this frequency 
band. In the first phase of this project, receiver systems consisting of 
V-dipole feeds and front-ends have been installed on four of the thirty
GMRT antennas. Test observations were carried out on a number of bright
3C sources. The initial results are encouraging. This paper will also
describe results of simultaneous observations carried out using the
existing GMRT correlator, the new GMRT software correlator and a system
employing digitization and direct recording of signals at two antenna
bases.
\end{abstract}

\vspace{-10mm}
\section{Introduction}
\label{s:introduction}

The Giant Metrewave Radio Telescope (GMRT) is an interferometric array
consisting of thirty 45-m diameter antennas spread over 25 km, operating
in frequency bands centred around 151, 325, 610/235, and 1400 MHz.
Receivers and feeds for operation at frequencies below 100 MHz are
currently not available.

A 50~MHz system for GMRT is being built by the Raman Research Institute, 
Bangalore, to carry out a low frequency survey of the sky visible to GMRT 
with arcmin resolution and better surface brightness sensitivity.
The survey results and the observing system will be made available to the 
astronomy community. With a primary beam area of 0.03 steradians, 
a mosaic of about 300 pointings is required to cover the northern sky. 
Expected survey parameters, at 50~MHz, assuming a 10~MHz bandwidth, 
are given below and a comparison with other major surveys is given in 
Table~\ref{t:surveycomparison}.
\begin{compactitem}
\item Nominal system temperature: $\approx 4000^{\circ}$K
(sky dominated).
\item Synthesized beam: $\sim 1'$.
\item RMS sensitivity (4~hr synthesis with dual polarization):
$\sim 2$~mJy~beam$^{-1}$ (thermal noise limited).
\end{compactitem}

This is about two orders of magnitude better than the 74~MHz
VLA survey. We expect the proposed 50~MHz survey to be at least an order
of magnitude better even if we do not achieve the thermal noise limited
sensitivity but get limited by the dynamic range.
\begin{table}[!t]
\tiny
\centering
\caption{\small Expected survey parameters and comparison with
major surveys.}
\vspace{2mm}
\begin{tabular}{r | c c c c c c}
\hline\hline \\
{\bf Survey} & {\bf Sky} & {\bf Telescope} & {\bf Surveying} & {\bf RMS}
& {\bf Equivalent RMS}\\
{\bf Name} & {\bf Coverage} & {\bf time } & {\bf speed} & {\bf noise}
& {\bf noise at 50 MHz}\\
(Frequency) & steradians & hours & hours/steradian & $\sim 1$-$\sigma$ in
mJy & $\sim 1$-$\sigma$ in mJy\\\\
\hline\hline\\
{\bf RRI-GMRT (50~MHz)} & 3$\pi$ & 1200 & 128 & 2(10)* & 2(10)*\\
{\bf NVSS (1.4~GHz)}    & 10.3   & 2700 & 262 & 0.45   & 6.5\\
{\bf WENSS (325~MHz)}   & $\pi$  & 960  & 305 & 3.6    & 16\\
{\bf VLSS (74~MHz)}     & $3\pi$ & 900  & 95  & 100    & 137\\\\
\hline
\end{tabular}
\\ * Terms in brackets denote dynamic range limited performance.
\label{t:surveycomparison}
\end{table}

\vspace{-3mm}
\section{Feed design}
The main goals of the feed system design were to obtain a reasonable
aperture efficiency ($0.5<\eta_{ap}<0.7$), symmetrical E~\&~H-plane
patterns and, a suitable physical dimension for co-locating it with
one of the existing feeds on the GMRT antenna turret, with minimum
interference to its operation. The desired frequency range of operation
(30-90~MHz) was chosen to facilitate imaging in the band protected for
radio astronomy (around 38~MHz), to have an overlap with the 74~MHz
system of VLA and to minimize the radio frequency interference (RFI) due
to the FM radio band starting beyond 90~MHz.

\begin{figure}[!t]
\centering
\subfigure[]{
\epsfig{figure=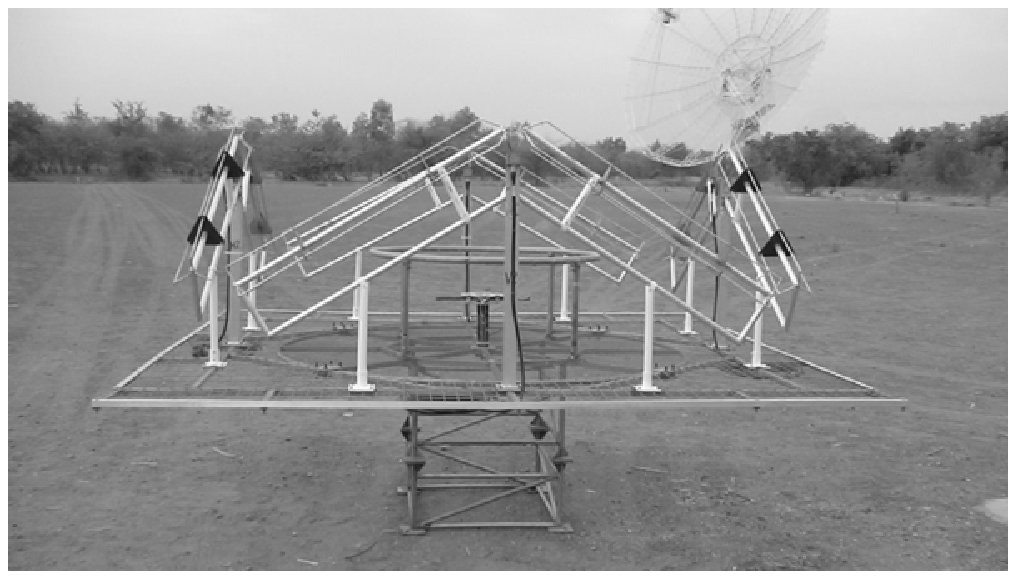,width=0.48\linewidth}}
\subfigure[]{
\epsfig{figure=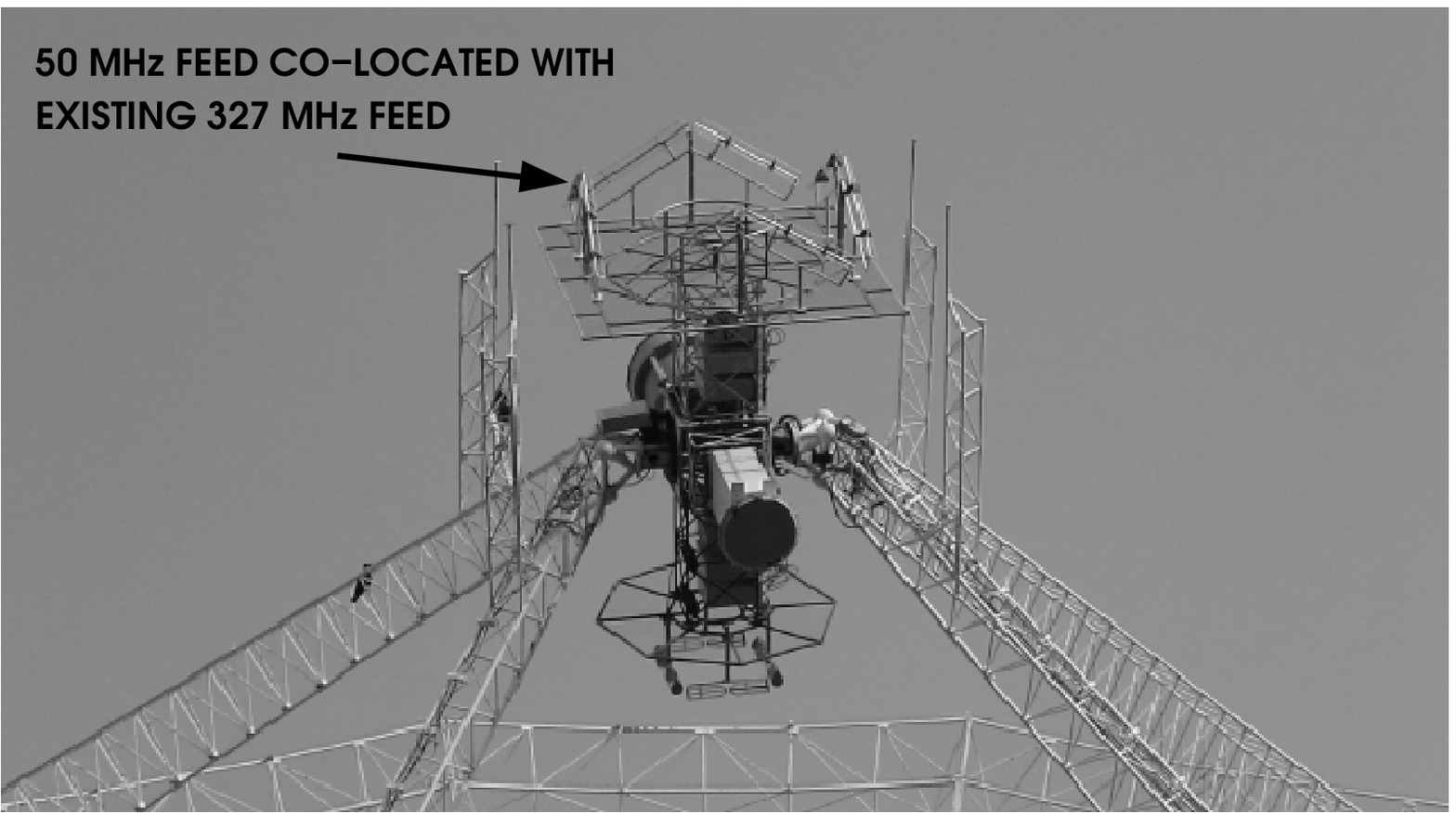,width=0.48\linewidth}}
\caption{\small (a) Four folded V-dipoles, each $\sim 2.4$~m length.
The feed point is positioned 1 m above a 3~m~$\times$~3~m square reflector
with a 50 mm mesh. (b) The feed co-located with the existing 327 MHz GMRT
feed. \normalsize}
\label{f:feeddesign}
\end{figure}

A feed system consisting of four folded V-dipoles in a {\it boxing-ring}
configuration (Fig.~\ref{f:feeddesign}a) was developed since it provides
a more symmetric E~\&~H-plane patterns than a single dipole feed
\citep{report:amiri07}. After extensive simulations and field trials, it
was decided to co-locate the new feed with the existing 327~MHz feed
(Fig.~\ref{f:feeddesign}b). A V-configuration of the dipole facilitated
accommodating the required length of the dipole despite the space
constraints. 
Four GMRT antennas, C04, C11, E02 and W02, were equipped with the new
V-dipole feeds. The maximum east-west baseline, E02-W02, is $\sim 6$~km,
while the north-south baseline, C04-C11, is $\sim 0.5$~km.

\vspace{-3mm}
\section{Test observations with the new feed system}
In September 2007, several 3C sources (including Cyg-A and Cas-A) were
observed using the existing GMRT receiver system. Observations of Cyg-A,
centred at 55~MHz, for an hour showed:
\begin{compactitem}
\item Aperture efficiency of $\sim 65$\%. This is based on the expected
background temperature given by \cite{mnras:turin73} and the flux
density of Cyg-A given by \cite{aanda:baars77}.
\item The rms noise observed in a single channel, with a bandwidth of
31.25~kHz, is 6\% of the flux density of Cyg-A.
\item The RMS of closure phase is $1.5^{\circ}$, which is as expected for
a measured SNR of 17 on the amplitude measurement of Cyg-A in each
baseline.
\end{compactitem}

Test observations carried out at GMRT showed that improvement in
signal-to-noise ratio (SNR) as a function of bandwidth ($\beta$) and
integration time ($\tau$) was not forthcoming when observed using the
existing GMRT receiver chain ~\citep{report:dwaraka08}. This is a well
known issue with GMRT continuum observations at other frequencies also.
The SNR saturated at a value of 17 for all the observed baselines. While
the reasons for this were not clear, RFI was suspected to be one of the
causes. To study if this limitation could be overcome by digitizing the
outputs of the low frequency front-end receivers right at antenna
bases, direct voltage recording systems were developed.

\vspace{-3mm}
\section{Direct voltage recording system}
\begin{figure}[!t]
\centering
\epsfig{figure=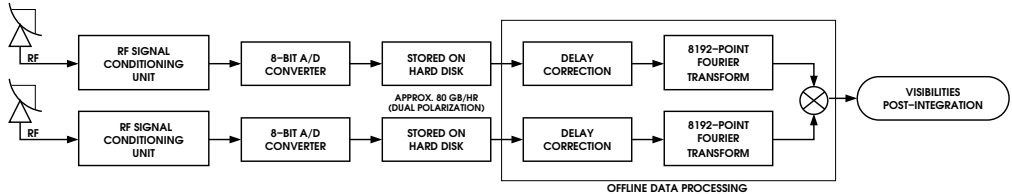,width=\linewidth}
\caption{\small Simplified block schematic of the direct
voltage recording system for a two-element interferometer. \normalsize }
\label{f:dvrblock}
\end{figure}

A simplified block schematic of the direct voltage recording system
developed at the Raman Research Institute (hereafter referred to as RRI-DS)
is shown in Fig.~\ref{f:dvrblock}. A detailed description of this system
can be found in \cite{thesis:girish09}.

RRI-DS digitizes a signal of $\sim 5.5$~MHz bandwidth using a sampling
clock of $\sim 11$~MHz. The digitized data is recorded on hard disks at a
rate of about 80~GB/hour (dual polarization). A GPS-disciplined rubidium
oscillator installed at each antenna base synchronizes the recording
systems with an accuracy of $\sim 5$~ns. The $\sim 1$~s averaged
visibilities are computed offline by averaging 1024, 8192-point FFT frames.

\vspace{-3mm}
\section{Test observations with a two-element interferometer}

\begin{figure}[!t]
\centering
\subfigure[]{
\epsfig{figure=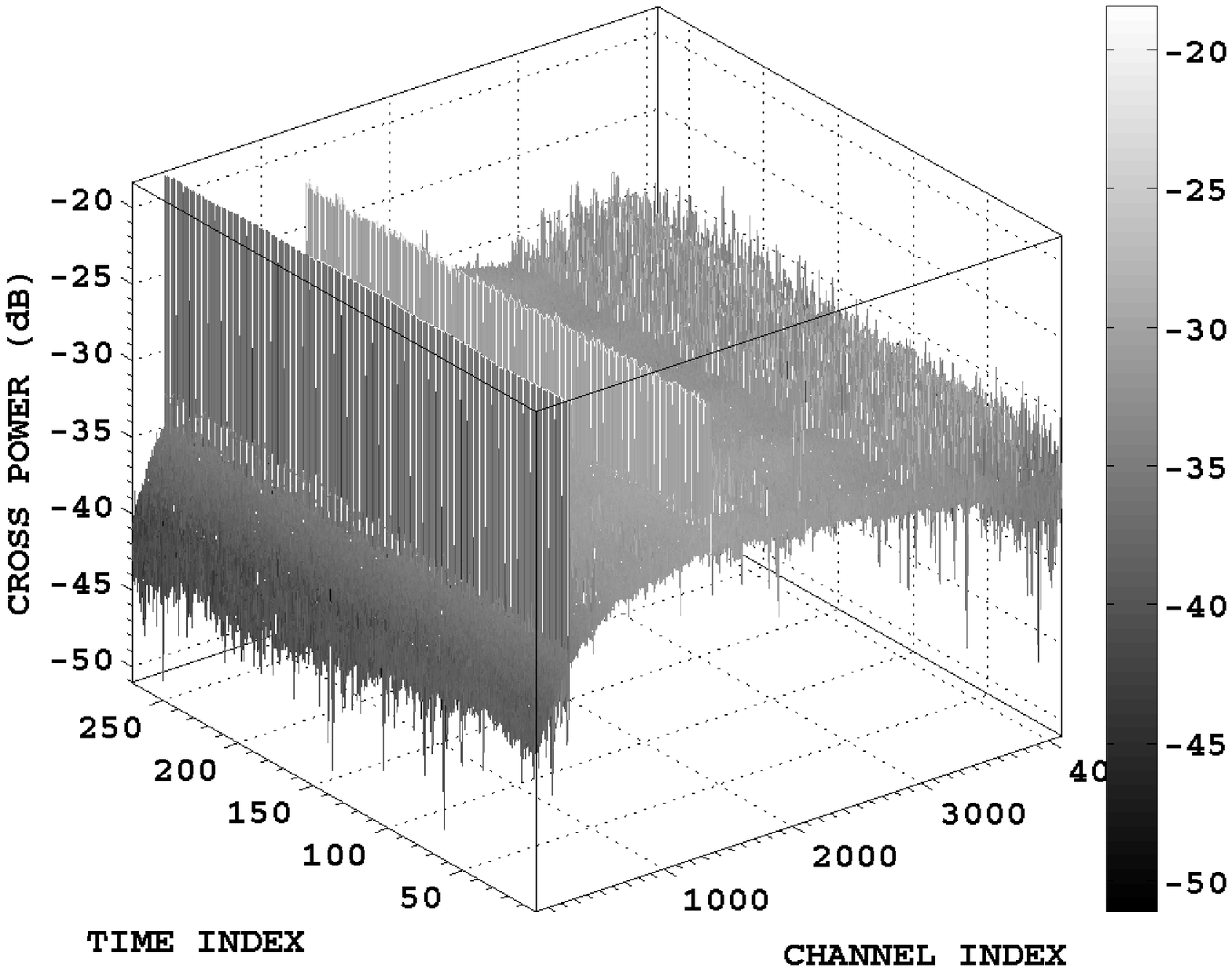,width=0.48\linewidth}}
\subfigure[]{
\epsfig{figure=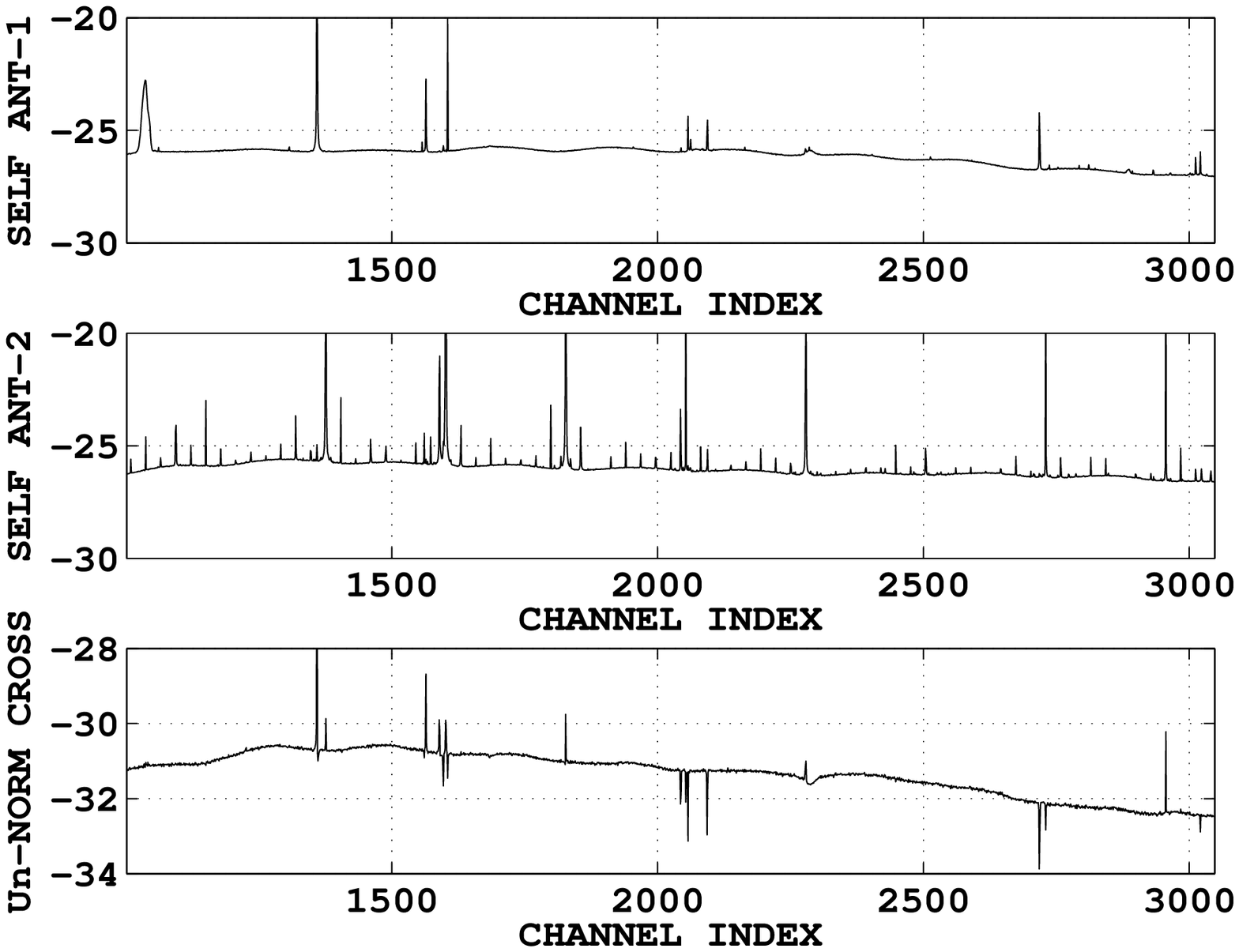,width=0.48\linewidth}}
\caption{ \small (a) Typical dynamic cross-power spectra of Cyg-A obtained
by RRI-DS; shown here for $\sim 4$~min (4096 frequency channels; $\sim 1$~s
averaged instantaneous time frame). The {\it x}-axis represents bandwidth
 of $\sim 5.5$~MHz centred at $\sim 55$~MHz, the $y$-axis represents
$\sim 4$~min observation time and the $z$-axis represents the cross-power
in dBs. (b)
Typical self-power and cross-power spectrum, for the central 2048 channels
(bandwidth $\sim 2.7$~MHz), obtained after 4~min integration.
\normalsize }
\label{f:selfcrosspectra}
\end{figure}

\begin{table}[!t]
\caption{Relevant specifications of the three systems.}
\centering
\tiny
\vspace{2mm}
\begin{tabular}{r | c c c} 
\hline\hline\\         
& {\bf GMRT-HC}  & {\bf GMRT-SC} & {\bf RRI-DS} \\\\
\hline 
\hline \\
{\bf Bandwidth (MHz)}            & 8     & 16    & 5.5 \\
{\bf Spectral Channels}          & 128   & 512   & 4096 \\
{\bf Spectral Resolution (kHz)}  & 62.5  & 31.25 & 1.36 \\
{\bf Integration time (s)}       & 1     & 1     & 0.75 \\\\
\hline
\end{tabular} 
\label{t:3systemspecs}
\end{table}

RRI-DSs were installed at antennas C04 and C11. In July 2008, simultaneous
test observations of 3C sources were carried out using GMRT hardware
correlator (GMRT-HC), GMRT software correlator (GMRT-SC) and RRI-DS, to
compare the performance of the three systems. For GMRT-HC and GMRT-SC,
the IF bandwidth was set to 6 MHz. Hence, only 3 MHz band was available in
each sideband. There are about 50 and 100 channels, within the 3~dB
bandwidth, in the GMRT-HC and GMRT-SC spectrum, respectively.
Table~\ref{t:3systemspecs} shows the relevant specifications of these
three systems.

\begin{figure}[!t]
\centering
\subfigure[]{
\epsfig{figure=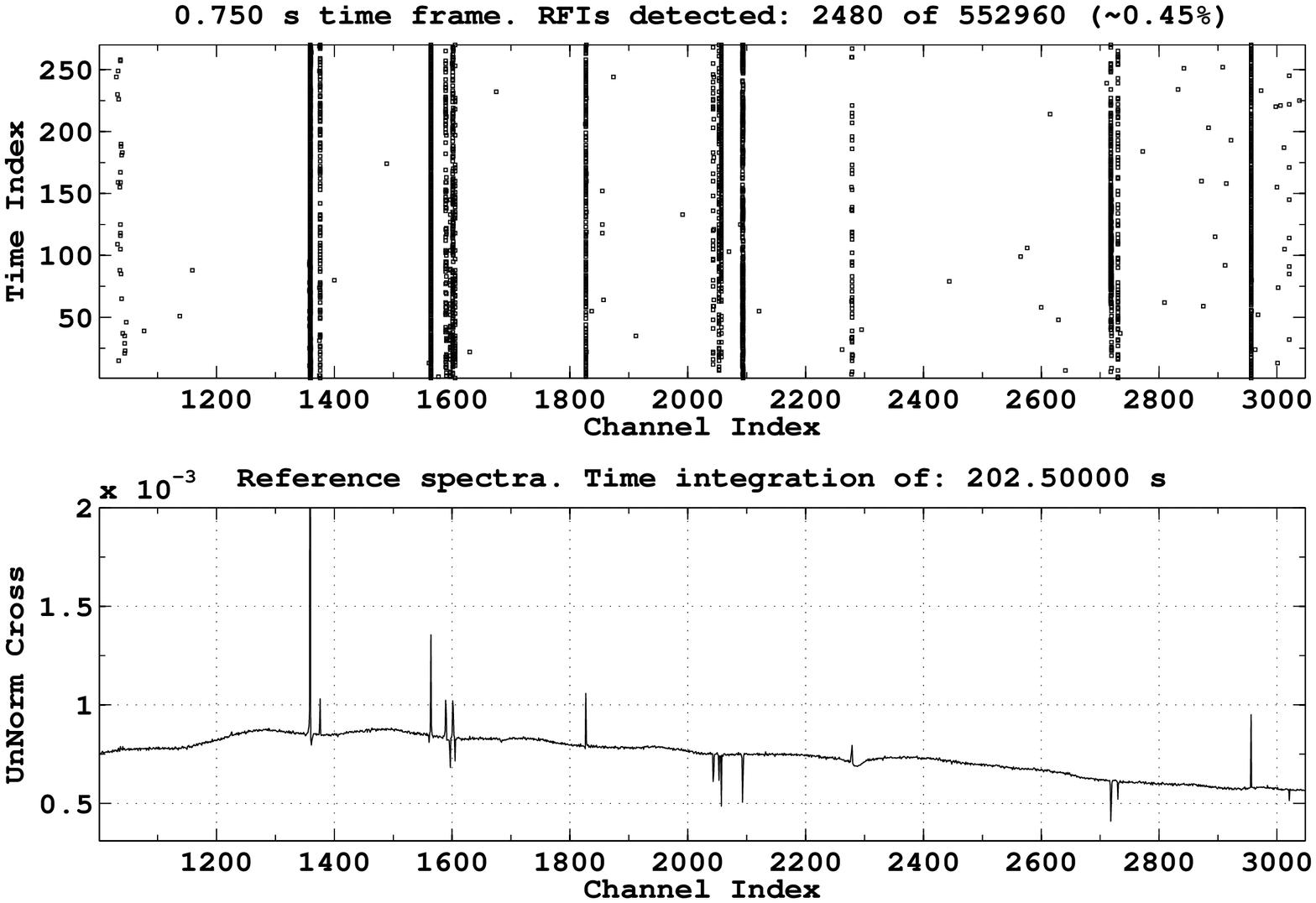,width=0.48\linewidth}}
\subfigure[]{
\epsfig{figure=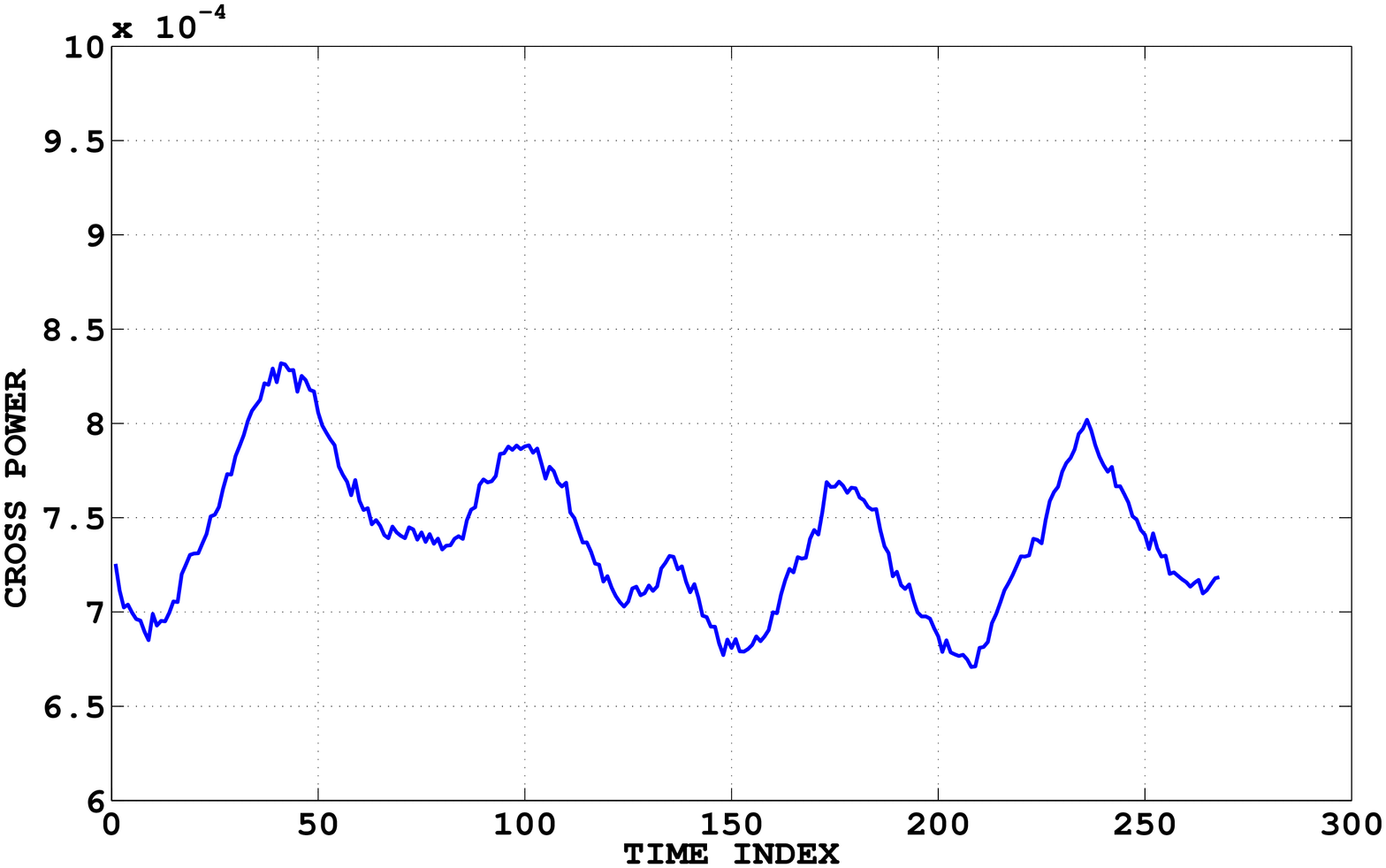,width=0.48\linewidth}}
\caption{ \small RFI scenario. (a) RFI points detected (above 4-$\sigma$)
for the dynamic spectra shown in Fig.~\ref{f:selfcrosspectra}a.
(b) A plot of variation in total power in the $\sim 1$~s averaged cross-power
spectra shown for about 4 min.
\normalsize }
\label{f:rfiscenario}
\end{figure}

Fig.~\ref{f:selfcrosspectra}a shows a dynamic 4096-point cross-power
spectra of Cyg-A for about 4~min with $\sim 1$~s integration time. Prior
to obtaining the cross-power spectra, we mitigate strong and short (spiked)
bursts of RFI in the temporal domain. Fig.~\ref{f:selfcrosspectra}b shows
central 2048 channels of self-power and cross-power spectrum (obtained with
4~min integration). Notice the {\it comb} in self-power spectrum of
antenna-2 (C11). Clearly, antenna C11 is affected more by RFI, compared to
antenna C04. From the dynamic spectra, each 1~s averaged instantaneous time
frame was band-normalized using a bandpass template. Any spectral channel
above the 4-$\sigma$ threshold was flagged as RFI affected channel.
Fig.~\ref{f:rfiscenario}a shows the RFI scenario for the 4 min data.
Statistics showed, $\sim 0.5$\% of the data was affected by RFI and there
were less than 10 interference points that were present continuously for
10~s or more. Fig.~\ref{f:rfiscenario}b shows a plot of variation of the
integrated cross-power in each of the 270, $\sim 1$~s averaged cross-power
spectra. It is clearly seen that there is a very low-frequency
{\it underlying time-structure} with a periodicity of about 45~s. This
underlying time-structure is seen at the output of all the three systems.

After a generous excision of RFI channels and removing the low-frequency
underlying structure, the SNR (mean/rms) measured after collapsing
channels and integrating in time is shown in Table~\ref{t:3systemsSNR}.
The SNR improvements measured in GMRT-SC and RRI-DS are better by factors
of about 2 and 3, respectively, when compared to GMRT-HC. However, the
absolute SNR values measured for RRI-DS are off by factor of 2-5,
compared to those expected.

\begin{table}[!t]
\centering
\tiny
\caption{SNR measured after removing the underlying time-structure.}
\vspace{2mm}
\begin{tabular}{r | c c c c c c c c c c}
\hline\hline\\
& \multicolumn{3}{c}{\bf GMRT-HC}  &
\multicolumn{3}{c}{\bf GMRT-SC}  &
\multicolumn{3}{c}{\bf RRI-DS} \\
{\bf Integration} & \multicolumn{3}{c}{\bf Channel BW (kHz)}  &
\multicolumn{3}{c}{\bf Channel BW (kHz)}  &
\multicolumn{3}{c}{\bf Channel BW (kHz)} \\
{\bf Time (s)} & 62.5 & 250 & 500 & 62.5 & 250 & 500
& 62.5 & 250 & 500\\\\
\hline
\hline \\
{\bf 1} & 41  & 43  & 55  &  66  &  80  &  105  &  87 (90)*   & 150 (180)*
&  175 (260)*\\
{\bf 4} & 43  & 44  & 55  &  75  &  83  &  114  &  129 (180)* & 187 (365)*
&  207 (520)*\\
{\bf 8} & 45  & 45  & 56  &  77  &  85  &  116  &  148 (260)* & 198 (520)*
&  215 (730)*\\\\
\hline
\end{tabular}
\\ * Numbers is round brackets show expected SNR.
\label{t:3systemsSNR}
\end{table}

\vspace{-3mm}
\section{Conclusions}
We have successfully designed, installed and tested a 50 MHz system on 4 
GMRT antennas. Aperture efficiency of $\sim 65$\% was measured in the
band centred at 55~MHz. Test observations indicate satisfactory
performance of the feed system. Investigations are underway to understand
the mitigation of systematics and RFI, to achieve performance limited by
thermal noise.

\small
\vspace{-3mm}
\section*{\small Acknowledgment}
We thank our colleagues at GMRT site, NCRA and, the Mechanical Engineering 
Services and the Radio Astronomy Lab of RRI, for their excellent support.


\end{document}